\documentclass[aps,prx,reprint,superscriptaddress,amssymb,amsmath]{revtex4-2}

\usepackage{bm}
\usepackage{braket}
\usepackage{mathrsfs}
\usepackage{graphicx}
\usepackage{glossaries}
\usepackage[colorlinks=true,allcolors=blue]{hyperref}
\usepackage[separate-uncertainty=true,range-units=single,range-phrase=--]{siunitx}

\newacronym{e-p}{$e$-ph}{electron-phonon}
\newacronym{XAS}{XAS}{x-ray absorption spectroscopy}
\newacronym{IXS}{IXS}{inelastic x-ray scattering}
\newacronym{RIXS}{RIXS}{resonant inelastic x-ray scattering}
\newacronym{ARPES}{ARPES}{angle-resolved photoemission spectroscopy}

\begin{document}

\title{Probing electron-phonon interactions away from the Fermi level with resonant inelastic x-ray scattering}

\author{C.~D.~Dashwood}\email{cameron.dashwood.17@ucl.ac.uk}
\thanks{These authors contributed equally}
\affiliation{London Centre for Nanotechnology and Department of Physics and Astronomy, University College London, London, WC1E 6BT, UK}

\author{A.~Geondzhian}
\thanks{These authors contributed equally}
\affiliation{Max Planck POSTECH/KOREA Research Initiative, 37673 Pohang, South Korea}
\affiliation{Max Planck Institute for the Structure and Dynamics of Matter, Luruper Chaussee 149, 22761 Hamburg, Germany}

\author{J.~G.~Vale}
\affiliation{London Centre for Nanotechnology and Department of Physics and Astronomy, University College London, London, WC1E 6BT, UK}

\author{A.~C.~Pakpour-Tabrizi}
\affiliation{London Centre for Nanotechnology and Department of Electronic and Electrical Engineering, University College London, London, WC1E 6BT, UK}

\author{C.~A.~Howard}
\author{Q.~Faure}
\author{L.~S.~I.~Veiga}
\affiliation{London Centre for Nanotechnology and Department of Physics and Astronomy, University College London, London, WC1E 6BT, UK}

\author{D.~Meyers}
\affiliation{Condensed Matter Physics and Materials Science Department, Brookhaven National Laboratory, Upton, New York 11973, USA}
\affiliation{Department of Physics, Oklahoma State University, Stillwater, Oklahoma 74078, USA}

\author{S.~G.~Chiuzb\u{a}ian}
\affiliation{Synchrotron SOLEIL, L'Orme des Merisiers, Saint-Aubin, BP 48, 91192 Gif-sur-Yvette, France}
\affiliation{Sorbonne Universit\'{e}, CNRS, Laboratoire de Chimie Physique-Mati\'{e}re et Rayonnement, UMR 7614, 4 place Jussieu, 75252 Paris Cedex 05, France}

\author{A.~Nicolaou}
\author{N.~Jaouen}
\affiliation{Synchrotron SOLEIL, L'Orme des Merisiers, Saint-Aubin, BP 48, 91192 Gif-sur-Yvette, France}

\author{R.~B.~Jackman}
\affiliation{London Centre for Nanotechnology and Department of Electronic and Electrical Engineering, University College London, London, WC1E 6BT, UK}

\author{A.~Nag}
\author{M.~Garc\'{i}a-Fern\'{a}ndez}
\author{Ke-Jin Zhou}
\author{A.~C.~Walters}
\affiliation{Diamond Light Source, Didcot, Oxfordshire, OX11 0DE, UK}

\author{K.~Gilmore}
\affiliation{Condensed Matter Physics and Materials Science Department, Brookhaven National Laboratory, Upton, New York 11973, USA}
\affiliation{Physics Department and IRIS Adlershof, Humboldt-Universit\"{a}t zu Berlin, Zum Gro{\ss}en Windkanal 2, 12489 Berlin, Germany}
\affiliation{European Theoretical Spectroscopy Facility (ETSF)}

\author{D.~F.~McMorrow}
\affiliation{London Centre for Nanotechnology and Department of Physics and Astronomy, University College London, London, WC1E 6BT, UK}

\author{M.~P.~M.~Dean}\email{mdean@bnl.gov}
\affiliation{Condensed Matter Physics and Materials Science Department, Brookhaven National Laboratory, Upton, New York 11973, USA}

\begin{abstract}
Interactions between electrons and lattice vibrations are responsible for a wide range of material properties and applications. Recently, there has been considerable interest in the development of \gls*{RIXS} as a tool for measuring \gls*{e-p} interactions. Here, we demonstrate the ability of \gls*{RIXS} to probe the interaction between phonons and specific electronic states both near to, and away from, the Fermi level. We performed carbon $K$-edge \gls*{RIXS} measurements on graphite, tuning the incident x-ray energy to separately probe the interactions of the $\pi^*$ and $\sigma^*$ electronic states. Our high-resolution data reveals detailed structure in the multi-phonon \gls*{RIXS} features that directly encodes the momentum dependence of the \gls*{e-p} interaction strength. We develop a Green's-function method to model this structure, which naturally accounts for the phonon and interaction-strength dispersions, as well as the mixing of phonon momenta in the intermediate state. This model shows that the differences between the spectra can be fully explained by contrasting trends of the \gls*{e-p} interaction through the Brillouin zone, being concentrated at the $\Gamma$ and $K$ points for the $\pi^*$ states, while being significant at all momenta for the $\sigma^*$ states. Our results advance the interpretation of phonon excitations in \gls*{RIXS}, and extend its applicability as a probe of \gls*{e-p} interactions to a new range of out-of-equilibrium situations.
\end{abstract}

\maketitle

\section{Introduction}

Knowledge of the interactions between electrons and phonons is central to understanding a diverse array of condensed matter phenomena. Basic material properties such as low-field electrical transport \cite{Gunnarsson2003}, as well as emergent collective phases including charge density waves \cite{Frohlich1954} and conventional superconductivity \cite{Bardeen1957}, are all crucially dependent on the coupling of phonons to electrons near the Fermi surface. Extensive experimental efforts have therefore been devoted to quantifying this coupling through various techniques. Neutron scattering \cite{Pintschovius2005}, non-resonant \gls*{IXS} \cite{Piscanec2004}, and Raman spectroscopy \cite{Ferrari2007} all probe \gls*{e-p} coupling via the lattice degrees of freedom, while \gls*{ARPES} \cite{Tanaka2013} measures it via the electronic self-energy. 

In other far-from-equilibrium situations, however, the interaction between phonons and electrons \textit{away} from the Fermi surface becomes important. This regime applies to processes such as high-temperature heat transport \cite{Zhou2020}, high-field electrical transport \cite{Yao2000}, and phonon-assisted optical transitions \cite{Novko2019, Caruso2021}, with applications in optoelectronics \cite{Wright2016, Handa2018}, for example. The established probes of \gls*{e-p} coupling, listed above, are generally unable to access such highly excited electronic states, particularly if information is required about the interaction with phonon modes of specific momenta.

Recently, \gls*{RIXS} has been gaining traction as a new technique to measure the \gls*{e-p} interaction strength, which manifests directly in the intensity of the phonon excitations. Over the last decade, improvements in energy resolution have enabled experimental studies on a growing number of materials \cite{Yavas2010, Lee2013, Lee2014, Moser2015, Fatale2016, Johnston2016, Meyers2018, Rossi2019, Vale2019, Braicovich2020, Geondzhian2020b, Li2020, Peng2020, Feng2020}, alongside an advancing theoretical understanding of the phonon contribution to the \gls*{RIXS} cross-section \cite{Ament2011, Devereaux2016, Geondzhian2018, Geondzhian2020a, Bieniasz2020}. Much of the excitement around these developments lies in the momentum-resolution that \gls*{RIXS} offers. The other great strength of \gls*{RIXS} -- the ability to probe a particular electronic orbital by tuning the incident energy to a specific absorption edge -- has thus far been under exploited.

Here, we demonstrate the power of \gls*{RIXS} to probe the interaction of phonons with two distinct electronic states in graphite: the $\pi^*$ state near the Fermi level, and the $\sigma^*$ state well above it. At the resonances of these two states, we find qualitatively different multi-phonon excitations in our \gls*{RIXS} spectra, showing a stark difference in the phonon momenta to which they couple. This behaviour cannot be captured by the currently available theoretical models, which assume dispersionless phonons \cite{Ament2011, Johnston2016, Bieniasz2020} or focus on one-phonon processes \cite{Devereaux2016}. We therefore extend a Green's-function approach, previously applied to small molecules \cite{Geondzhian2018}, to treat the full momentum-dependence of phonons in a crystalline lattice. This model accurately reproduces our experimental spectra at both resonances, revealing contrasting trends of the \gls*{e-p} interaction strength through the Brillouin zone.

\begin{figure}
	\includegraphics[width=\linewidth]{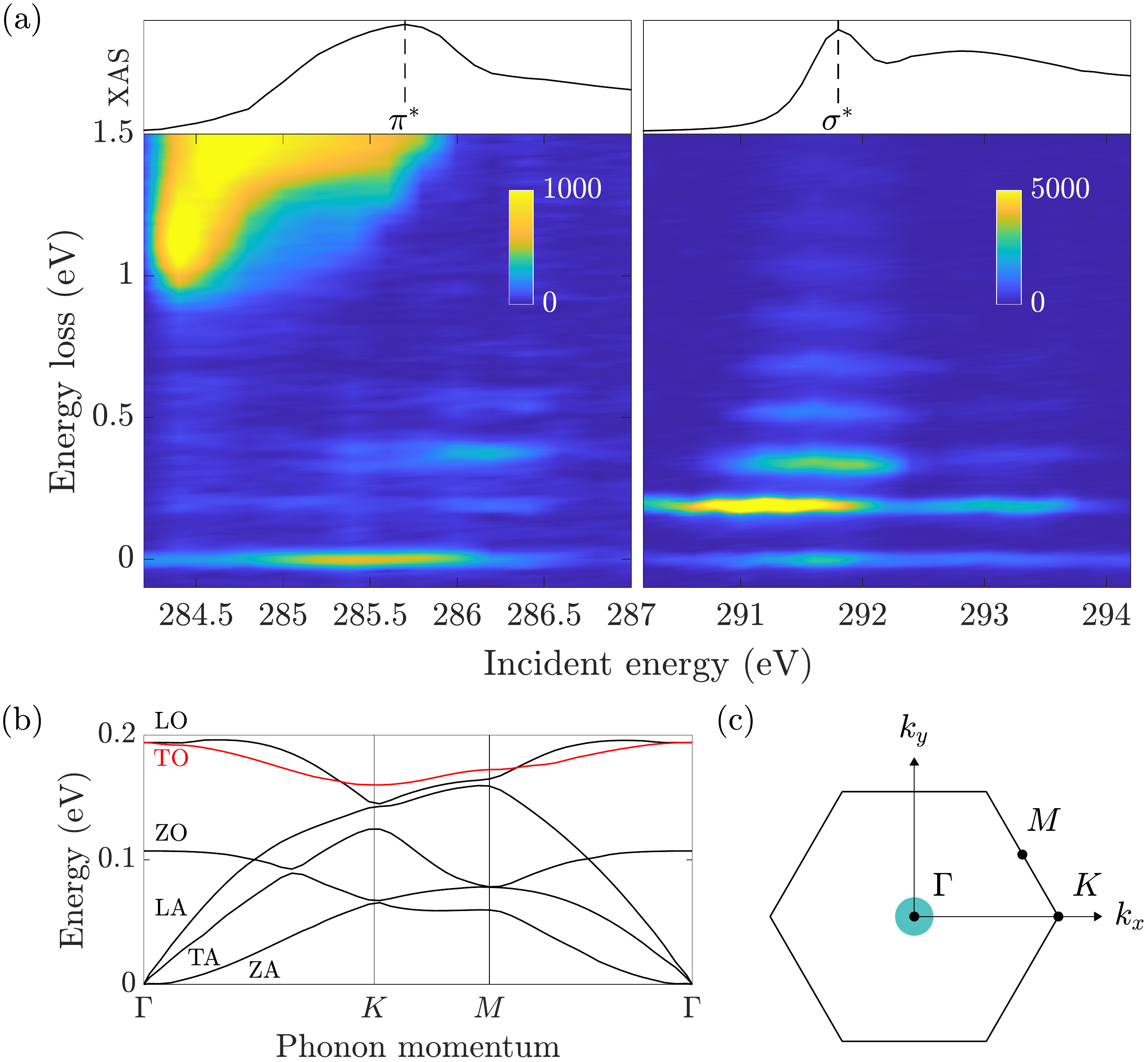}
	\caption{\label{RIXS_maps}Contrasting \gls*{RIXS} response at the $\pi^*$ and $\sigma^*$ resonances. (a) \gls*{RIXS} maps around both resonances, with \gls*{XAS} above for reference (the dashed vertical lines in the \gls*{XAS} mark the peaks of the resonances at \SI{285.6}{\electronvolt} and \SI{298.1}{\electronvolt} respectively). Both maps show a series of phonon features above the elastic line, with the contrasting resonance behavior most apparent for the two-phonon feature between \SIrange{0.3}{0.4}{\electronvolt}. The intense feature above $\sim$\SI{1}{\electronvolt} in the $\pi^*$ map arises from electronic transitions. (b) Phonon dispersion of graphite from Ref.~\cite{Mohr2007}, with the transverse (T), longitudinal (L) and out-of-plane (Z) acoustic (A) and optical (O) modes marked. The TO mode with significant coupling is marked in red. (c) The 2D-projected Brillouin zone of graphite, with the momenta accessible in our \gls*{RIXS} measurement indicated by the turquoise circle, and high symmetry positions $\Gamma$, $K$, and $M$ marked.}
\end{figure}

\section{RIXS measurements on graphite}

Graphite is a favorable material for this study, as it has a simple electronic and phononic structure, as well as two well-defined excitations to the low-energy $\pi^*$ and high-energy $\sigma^*$ states at the carbon $K$ edge. Coupling to the electronic states near the Fermi level is known to be dominated by the highest optical phonon band above \SI{0.15}{\electronvolt} \cite{Piscanec2004, Lazzeri2006, Ferrari2007, Park2008, Na2019}, which is well resolved by our \SI{47}{\milli\electronvolt}-resolution measurements. There are also no other low-energy excitations (such as magnons) that could obscure the phonon excitations in our \gls*{RIXS} spectra. This allows us to validate our methodology at the $\pi^*$ resonance, before extending our treatment to the $\sigma^*$ states. Given their similarities to graphite \cite{Lazzeri2006, Ferrari2007, Park2008}, our results are also applicable to the other technologically-important allotropes of carbon.

Our carbon $K$-edge \gls*{RIXS} measurements on graphite are summarized in Fig.~\ref{RIXS_maps}(a), which shows approximately zone-center \gls*{RIXS} maps around the $\pi^*$ and $\sigma^*$ resonances (see Appendix \ref{Exp_details} for experimental details). Above the elastic line at \SI{0}{\electronvolt} energy loss, both maps show a series of features of decreasing intensity that are reminiscent of the harmonic progression of phonon excitations seen in previous studies \cite{Lee2013, Lee2014, Moser2015, Fatale2016, Johnston2016, Meyers2018, Rossi2019, Vale2019, Braicovich2020, Geondzhian2020b, Li2020, Peng2020, Feng2020}. These features are more intense in the $\sigma^*$ map, suggesting a stronger overall coupling.

\begin{figure*}
	\includegraphics[width=\linewidth]{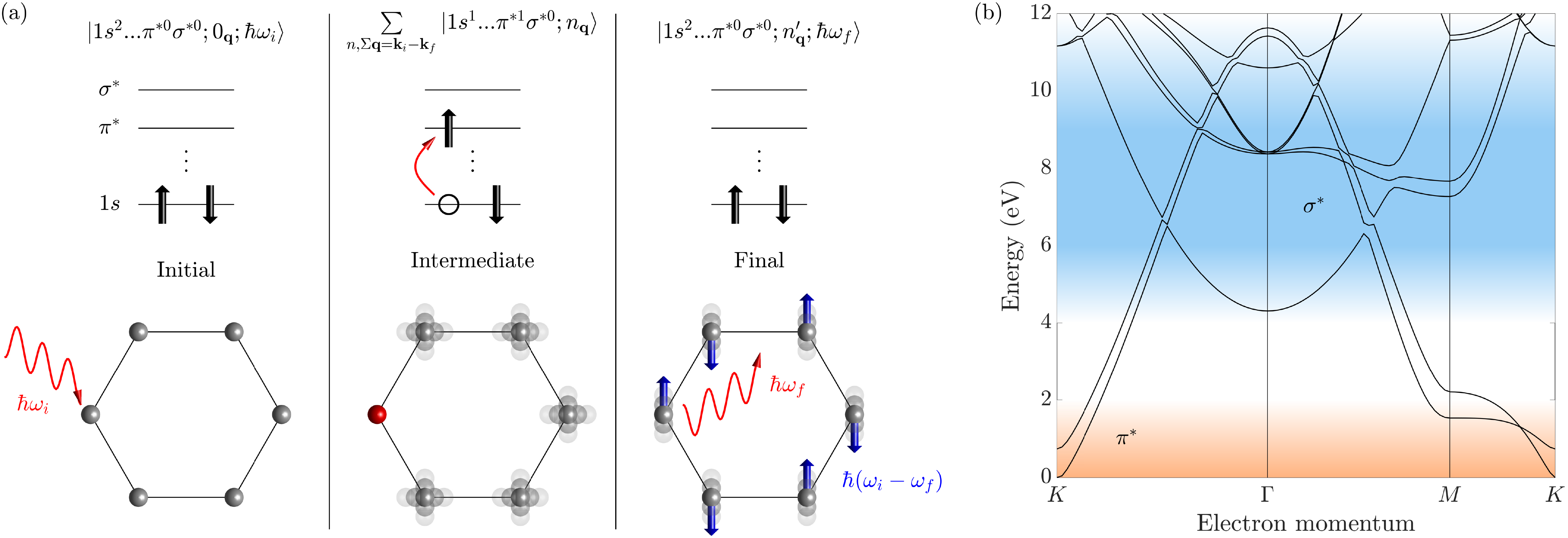}
	\caption{\label{RIXS_process}Phonon generation in \gls*{RIXS}. (a) Schematic of the \gls*{RIXS} process at the carbon $K$ edge. An incident photon of energy $\hbar\omega_i$ and wavevector $\bm{k}_i$ excites an electron from the $1s$ to the $\pi^*$ or $\sigma^*$ bands, leaving a core hole. In the intermediate state, the altered charge density perturbs the positions of surrounding ions and generates $n_{\bm{q}}$ phonon modes. The excited electron then relaxes via the emission of a photon of energy $\hbar\omega_f$, leaving $n_{\bm{q}}'$ phonons with total energy $\hbar(\omega_i - \omega_f)$ in the final state (the zone-center TO mode is depicted \cite{Mohr2007}). (b) Electronic band structure of graphite from Ref.~\cite{Marinopoulos2004}, with the low-energy $\pi^*$ bands shaded in orange and the high-energy $\sigma^*$ bands in blue.}
\end{figure*}

\begin{figure*}
	\includegraphics[width=\linewidth]{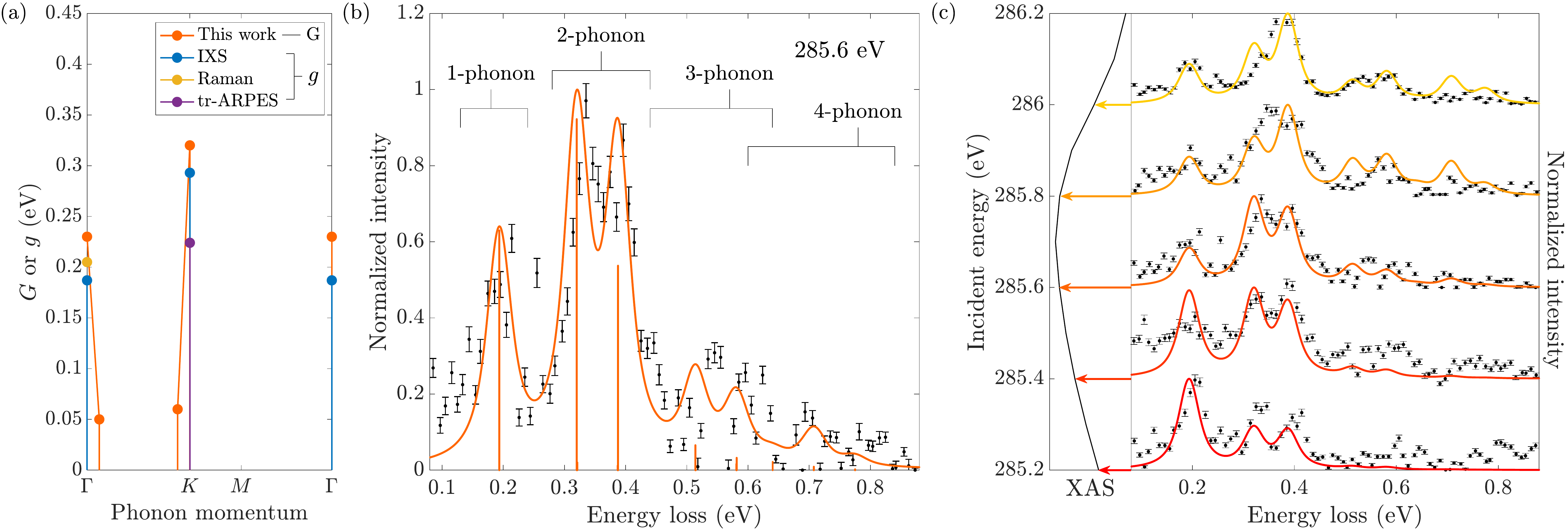}
	\caption{\label{pi_resonance}\gls*{e-p} interactions for low-energy $\pi^*$ excitations. (a) Modeled momentum dependence of $G$ through the 2D Brillouin zone (orange), compared to $g$ determined by \gls*{IXS} (blue) \cite{Piscanec2004}, Raman spectroscopy (yellow) \cite{Ferrari2007} and time-resolved \gls*{ARPES} (purple) \cite{Na2019}. (b) Normalized experimental (black points with error bars) and calculated (orange line, including experimental broadening) \gls*{RIXS} spectra on resonance (\SI{285.6}{\electronvolt}). The contribution from each phonon generation process is indicated by the vertical orange lines, grouped above by the number of phonons. (c) Incident-energy dependence of the experimental (black points with error bars) and calculated (red to orange lines) spectra, plotted alongside the \gls*{XAS} with arrows indicating the energies.}
\end{figure*}

We assign contributions to the phonon features through reference to the known phonon dispersion of graphite, shown in Fig.~\ref{RIXS_maps}(b) \cite{Mohr2007}. The lowest-energy feature occurs at $\sim$\SI{0.19}{\electronvolt} in both maps, which corresponds to the degenerate transverse (TO) and longitudinal optical (LO) modes at the $\Gamma$ point [whose displacement pattern is shown in the final state in Fig.~\ref{RIXS_process}(a)]. In a purely local picture, successive peaks would then correspond to higher harmonics of this first peak. The next feature around \SIrange{0.3}{0.4}{\electronvolt} energy loss, however, shows an internal structure that differs markedly between the two maps. At the $\pi^*$ resonance (\SI{285.6}{\electronvolt}) the feature is split into two (as can also be seen in the linecuts in Fig.~\ref{pi_resonance}), while at the $\sigma^*$ resonance (\SI{291.8}{\electronvolt}) a single broader asymmetric peak is visible.

To conserve momentum, a single-phonon excitation in a zone-centre spectrum can include only a zone-centre mode in the final state. The final states of a two-phonon feature, however, can conceivably consist of pairs of opposite-momenta phonons at any point in the Brillouin zone. The split feature at the $\pi^*$ resonance has peak energies of \SI{0.32}{\electronvolt} and \SI{0.39}{\electronvolt}, which correspond to double the energies of the TO mode at $K$ and $\Gamma$ respectively (see the Supplemental Material). Meanwhile, the broad feature at the $\sigma^*$ resonance spans double the full bandwidth of the TO mode. Phonons of different momenta therefore appear to be excited at each of the resonances.

\section{Electron-phonon interactions in RIXS}

We have identified the modes that contribute to the \gls*{RIXS} spectra, but why do phonons of different momenta appear at the $\pi^*$ and $\sigma^*$ resonances? To answer this, we need to understand how phonons are generated in a \gls*{RIXS} measurement.

Figure \ref{RIXS_process}(a) shows a schematic of the \gls*{RIXS} process at the carbon $K$ edge. An incident x-ray photon excites an electron from the $1s$ to the $\pi^*$ or $\sigma^*$ bands, depending on the energy, where it experiences a partially-screened potential from the resulting core hole. The altered charge density of this intermediate state perturbs the positions of the surrounding ions, generating multiple phonons. Finally, the excited electron relaxes to fill the core hole through the emission of a photon, with the difference in incident and scattered photon energies equal to the energy of the phonons remaining in the final state.

The composition of the intermediate electronic state can therefore have a significant impact on the phonon modes that are excited. As both resonances involve the same core-hole, this cannot explain the difference between the \gls*{RIXS} spectra. We instead focus on the band structure of the excited electron, shown in Fig.~\ref{RIXS_process}(b). For electrons excited into the low-energy $\pi^*$ states, the steepness of the bands confines them to a small region around the Dirac points at $K$. These states can only interact with phonons with momenta close to $\Gamma$, which scatter electrons within a single Dirac cone, or $K$, which scatter electrons between different Dirac cones. By contrast, the high-energy $\sigma^*$ bands are flatter and electrons excited into them can span the Brillouin zone, suggesting interactions with a wide range of phonon momenta. This provides an intuitive understanding of our earlier assignment of the \gls*{RIXS} features, where the split two-phonon feature at the $\pi^*$ resonance is dominated by phonons at $\Gamma$ and $K$, while the single broad feature at the $\sigma^*$ resonance has contributions from across the TO mode dispersion.

To put this on a more rigorous footing, we need a theoretical model that can account for interactions of the full excitonic intermediate state with multiple phonons of different momenta. Within the exact-diagonalization approaches proposed thus far \cite{Ament2011, Johnston2016, Devereaux2016}, this requires a computationally-intractable expansion of the Hilbert space. We therefore employed a Green's-function approach which implicitly accounts for vibronic effects in the intermediate state (full details can be found in Appendix \ref{Theoretical_model}). We address \gls*{e-p} interactions within the framework of a linked-cluster (cumulant) expansion using a second-order Fan-Migdal-type self-energy in the time domain. \gls*{RIXS} cross-sections were obtained by considering the non-diagonal elements of an exciton Green's function, as described elsewhere \cite{Geondzhian2018}. As well as fully accounting for the phonon dispersion, this approach allows a proper treatment of mode mixing in the intermediate state, which leads to significant renormalization of the spectral features.

In order to focus on the effect of the interaction-strength dispersion, we treat both $\pi^*$ and $\sigma^*$ intermediate states as localised excitons, which is a good approximation over the intermediate-state lifetime of graphite \cite{Bruhwiler1995}. The intermediate-state lifetime itself was determined using the core-hole lifetime from x-ray photoemission spectroscopy \cite{Sette1990}, adjusted to mimic the presence of the extra electron (see the Supplemental Material). Given the weak inter-layer coupling of graphite, the optical phonon modes of interest have a minimal dependence on the out-of-plane momentum \cite{Mohr2007} and we can safely restrict our analysis to the 2D-projected Brillouin zone. We used the phonon dispersion calculated with density functional perturbation theory, which is in good agreement with \gls*{IXS} measurements \cite{Mohr2007}. The only free parameter of the model is then the momentum-dependent \gls*{e-p} interaction strength, $G(\bm{q})$, which can be determined by fitting the experimental \gls*{RIXS} spectra.

\section{Electron-phonon interactions near the Fermi level}

We turn first to our data around the $\pi^*$ resonance. From our earlier assignment of the \gls*{RIXS} features, it appears that significant contributions come only from the TO mode at the $K$ point, and degenerate TO/LO modes at the $\Gamma$ point (see the Supplemental Material). We therefore include only the TO mode dispersion in our modelling, with a minimal set of momentum points: the $\Gamma$ and $K$ points, the experimental in-plane momentum transfer $\bm{q}_{\text{exp}} \approx (0.08, 0.00)$ r.l.u., and $K - \bm{q}_{\text{exp}}$. Inclusion of the latter two effectively captures the width over which $G(\bm{q})$ falls off away from the high-symmetry points. The intensity of the one-phonon peak is sensitive to the fact that $\bm{q}_{\text{exp}} \neq \Gamma$, but for the multi-phonon features it is sufficient to approximate the total momentum transfer as zero.

Figure \ref{pi_resonance}(b) shows a fit of this model to the spectrum on resonance, with the corresponding momentum dependence of $G(\bm{q})$ plotted in Fig.~\ref{pi_resonance}(a). We see good agreement between the experimental and calculated spectra, including the striking split two-phonon feature, despite our simple parameterization of $G(\bm{q})$. This confirms that the splitting arises due to the dominant contributions from pairs of phonons at $K$ and $\Gamma$, indicated by the vertical lines at \SI{0.32}{\electronvolt} and \SI{0.39}{\electronvolt} respectively. Intensity above \SI{0.45}{\electronvolt} is attributed to overlapping three- and four-phonon features, which are also consistent with contributions coming from pairs of $\pm K$-momenta phonons with additional $\Gamma$ modes.

We further tested our model through its ability to reproduce spectra with the incident x-ray energy detuned from the maximum of the \gls*{XAS}. Figure \ref{pi_resonance}(c) shows the energy dependence of the experimental and calculated spectra through the resonance. As well as the persistent splitting of the two-phonon feature below the resonance, our model captures the changing relative intensities of the one- and two-phonon features. The non-monotonic energy dependence of these intensities, including the offset of the maximum of the one-phonon peak intensity from zero-detuning \cite{Feng2020}, arises from a combination of strong \gls*{e-p} interaction strength and long core-hole lifetime \cite{Geondzhian2020a}. We emphasise that the only parameter being varied between the spectra in Fig.~\ref{pi_resonance}(c) is the incident energy, providing a strong validation of our model.

As the incident energy is raised through the upper tail of the \gls*{XAS} peak, the composition of the intermediate electronic state will change (picking up contributions from the $M$ point, see Fig.~\ref{RIXS_process}(b)] \cite{Olovsson2019}. This in turn allows scattering by phonons of different momenta, changing the momentum-dependence of $G$ and explaining the discrepancies between the experimental and calculated spectra above \SI{285.6}{\electronvolt}. Agreement can be improved by allowing $G(\bm{q})$ to vary slightly with incident energy (see the Supplemental Material).

We can now compare our results with those from other techniques. Alongside our best-fit $G(\bm{q})$, Fig.~\ref{pi_resonance}(a) also shows the \gls*{e-p} interaction strength determined by \gls*{IXS} \cite{Piscanec2004} and Raman spectroscopy \cite{Ferrari2007}, as well as a novel time-resolved \gls*{ARPES} technique \cite{Na2019}. We see reasonable agreement between all the techniques. Note, however, that we use a lower-case $g$ for the coupling probed by these other techniques, to highlight that they do not involve a core hole. As discussed above, \gls*{RIXS} measures a related but inequivalent coupling, $G$, of phonons to an excitonic intermediate state consisting of both the excited electron and core hole \cite{Geondzhian2018}. The concept of \textit{exciton}-phonon coupling is not new, being relevant to Raman and other optical spectroscopies \cite{Jiang2007, Antonius2017} (the only difference here being that the hole is often in the valence band rather than core level). Recent work in these fields has shown that, under certain approximations, the two quantities $G$ and $g$ can be related by \cite{Chen2020}
\begin{widetext}
    \begin{equation}
        G_{nm}^\nu(\bm{Q}, \bm{q}) = \sum_{\bm{k}} \left[\sum_{h,c,c'} A_{h(\bm{k}), c(\bm{k} + \bm{Q} + \bm{q})}^{m(\bm{Q} + \bm{q}) *} A_{h(\bm{k}), c'(\bm{k} + \bm{Q})}^{n(\bm{Q})} g_{c'c}^\nu(\bm{k} + \bm{Q}, \bm{q}) - \sum_{c,h,h'} A_{h(\bm{k} - \bm{q}), c(\bm{k} + \bm{Q})}^{m(\bm{Q} + \bm{q}) *} A_{h'(\bm{k}), c(\bm{k} + \bm{Q})}^{n(\bm{Q})} g_{hh'}^\nu(\bm{k} - \bm{Q}, \bm{q})\right]. \label{RIXS_coupling}
    \end{equation}
\end{widetext}
This expression describes an exciton in initial state $n$, with centre-of-mass momentum $\bm{Q}$, and wave-function $\ket{S_n} = \sum_{hc} A_{hc}^{n} \ket{hc}$, where $h$ ($c$) indicate hole (conduction) states. The exciton is scattered by a phonon in mode $\nu$ with momentum $\bm{q}$, to a final state $m$ with momentum $\bm{Q} + \bm{q}$ and wave-function $\ket{S_m} = \sum_{h'c'} A_{h'c'}^{m} \ket{h'c'}$. It can be seen that Eq.~\ref{RIXS_coupling} is separated into two terms, the first accounting for electron-phonon scattering with the hole-state unaffected, and the second for hole-phonon scattering with a passive electron, each weighted by the coefficients of the excitonic wavefunction. In cases where the core hole is strongly screened, \gls*{RIXS} reflects the single-particle coupling of the excited electron. When the core hole is weakly screened, \gls*{RIXS} accesses exciton-phonon coupling, which is the quantity of interest for the many out-of-equilibrium situations discussed in the Introduction.

From Fig.~\ref{pi_resonance}(a), we can also see how the various techniques probe different components of $g_{cc'}^\nu(\bm{Q}, \bm{q})$ and $G_{nm}^\nu(\bm{Q}, \bm{q})$. Standard \gls*{ARPES} resolves $g$ by electronic momentum $\bm{Q}$ for bands $c$ below the Fermi energy, but integrates over phonon modes $\nu$ and momenta $\bm{q}$ \cite{Tanaka2013}. By contrast, \gls*{IXS} \cite{Piscanec2004} and neutron scattering \cite{Pintschovius2005} resolve the phonon momentum $\bm{q}$  throughout the Brillouin zone for all modes $\nu$, but average over the equilibrium electronic state. Raman spectroscopy \cite{Ferrari2007} and the novel time-resolved \gls*{ARPES} \cite{Na2019} technique can access low-energy electronic excitations, but are restricted to particular phonon momenta ($\bm{q} = \Gamma$ and $K$ for graphite). By contrast, we have shown that \gls*{RIXS} can resolve $G$ by electronic state $n$, phonon mode $\nu$ and phonon momentum $\bm{q}$, integrating only over the electronic momentum $\bm{Q}$.

By modelling the contributions to the multi-phonon features, we have therefore determined the interactions between the $\pi^*$ electronic state and TO phonons throughout the Brillouin zone, all from a single zone-center spectrum. This method will be invaluable, as the soft x-ray edges at which phonons are enhanced (due to the long intermediate-state lifetimes) and energy resolutions are maximised, also suffer from restricted momentum transfer [see Fig.~\ref{RIXS_maps}(c)].

\begin{figure*}
	\includegraphics[width=\linewidth]{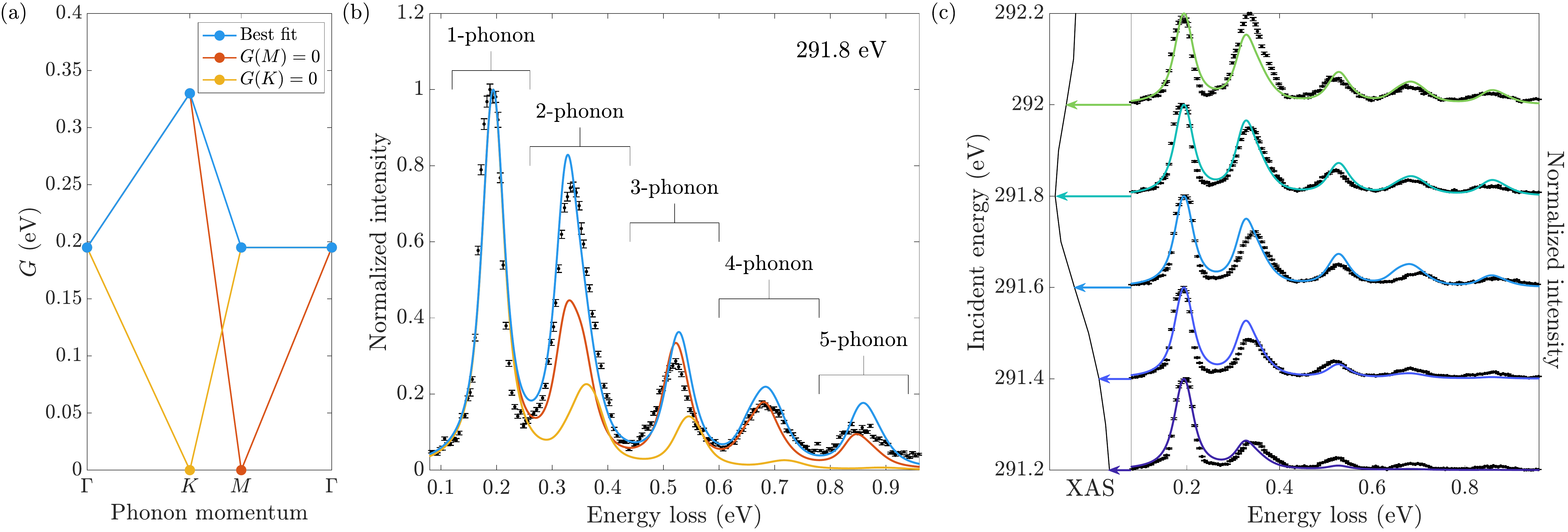}
	\caption{\label{sigma_resonance}\gls*{e-p} interactions for high-energy $\sigma^*$ excitations. (a) Best-fit momentum dependence of $G$ (blue), and that with $G(M) = 0$ (orange), and $G(K) = 0$ (yellow). (b) Normalized experimental (black points with error bars) and calculated [blue, orange, and yellow lines, corresponding to those in (a)] \gls*{RIXS} spectra on resonance (\SI{291.8}{\electronvolt}). The contribution from each $n$-phonon process is labeled above. (c) Incident-energy dependence of the experimental (black points with error bars) and calculated (blue to green lines) spectra, plotted alongside the \gls*{XAS} with arrows indicating the energies.}
\end{figure*}

\section{Electron-phonon interactions away from the Fermi level}

Having successfully reproduced the $\pi^*$ spectra, we now come to the high-energy $\sigma^*$ states. Based on our assignment of the \gls*{RIXS} features, we again restrict our analysis to the TO mode (see the Supplemental Material). In this case, however, the broad profile of the two-phonon peak suggests $G(\bm{q})$ is finite over a wide range of momenta. In order to constrain our parameter space, we therefore assume a linear dependence between the high-symmetry points $\Gamma$, $K$, and $M$. Given the more uniform variation of $G(\bm{q})$ around $\Gamma$, we can also approximate the experimental momentum transfer as zero for the whole spectrum.

Figure \ref{sigma_resonance}(a) shows the best-fit dispersion of $G(\bm{q})$, which we find to be significant at all the high-symmetry points. The resulting spectrum on resonance is shown in Fig.~\ref{sigma_resonance}(b), and again we see that a simple parameterization of reciprocal space is able to describe the experimental data well. Unlike the splitting seen at the $\pi^*$ resonance, here our model reproduces the subtle broadening and asymmetry of the two-phonon feature, confirming that these characteristics arise from the varying contributions of pairs of opposite-momenta phonons which disperse gradually through the Brillouin zone. 

In order to determine the sensitivity of our fit to the values at the high-symmetry points, we also show spectra in Fig.~\ref{sigma_resonance}(b) calculated with $G(M) = 0$ and $G(K) = 0$ ($G(\Gamma)$ must be finite to produce a significant one-phonon peak). We can see that the calculated spectra are most sensitive to the value of $G(K)$, which strongly influences the intensities of all the multi-phonon peaks. While the value of $G(M)$ has little impact on the three-, four- and five-phonon peaks, we see that a finite value on the order of $G(\Gamma)$ is required to accurately capture the intensity and shape of the two-phonon peak. We also show in the Supplemental Material how a previous \gls*{RIXS} measurement of $G(\bm{q})$ at the $\sigma^*$ resonance of graphite \cite{Feng2020}, which assumed zero coupling between the high-symmetry points, is unable to reproduce the multi-phonon features.

Our best-fit $G(\bm{q})$ is also able to reproduce the changing intensities of the features on detuning below \SI{291.8}{\electronvolt}, as shown in Fig.~\ref{sigma_resonance}(c). As for the $\pi^*$ states, we expect agreement between the model and experiment to degrade above the resonance due to changes in the intermediate electronic state. Further, just above the $\sigma_1^*$ resonance, that we are focused on, lies the $\sigma_2^*$ resonance [see Fig.~\ref{RIXS_process}(b)], which arises from zero-point vibrations \cite{Olovsson2019} that are not included in our modelling.

\section{Conclusion}

We have demonstrated that, through careful choice of the incident photon energy, \gls*{RIXS} is able to probe the interactions between phonons and electrons both close to, and away from, the Fermi level. Our results for graphite reveal that the momentum dependence of the interaction strength, $G(\bm{q})$, can be highly distinct in these two regimes: for low-energy $\pi^*$ states it is concentrated in small regions around the $\Gamma$ and $K$ points, while for high-energy $\sigma^*$ states it is significant across the Brillouin zone. This establishes \gls*{RIXS} as a uniquely versatile probe of \gls*{e-p} interactions in crystalline materials, applicable to studies of out-of-equilibrium processes alongside ordered phases in correlated systems. Combined with the recent development of ultrafast time-resolved \gls*{RIXS} at x-ray free-electron lasers \cite{Dean2016, Cao2019, Mitrano2020, Mazzone2021}, our methodology should prove particularly powerful for determining \gls*{e-p} interactions in pumped systems.

We find that contributions to the multi-phonon features in a \gls*{RIXS} spectrum can come from phonons away from the experimental momentum transfer, allowing us to constrain $G(\bm{q})$ throughout the Brillouin zone from a single zone-center spectrum. This is particularly apparent at the $\pi^*$ resonance, where the rapid variation of $G$ with momentum produces a split two-phonon feature. These contributions are also important, however, for the $\sigma^*$ states with more uniform $G(\bm{q})$. Here, the gradually dispersing phonons cause a broadening and asymmetry of the multi-phonon features, as has been seen in other studies \cite{Lee2013, Vale2019}, alongside shifts in their peak energies that could be mistaken for anharmonic effects \cite{Vale2019}. As we show in Appendix \ref{Theoretical_model}, mode mixing in the intermediate state results in the multi-phonon contributions affecting the intensity and resonance behavior of \textit{all} the phonon features in the spectra, including the one-phonon peak. Neglecting this effect is likely to lead to qualitative errors in the determination of $G(\bm{q})$ \cite{Feng2020}. These insights from graphite will be important when analysing the phonon excitations in RIXS measurements of more complex materials, such as the much-studied cuprates \cite{Yavas2010, Lee2013, Lee2014, Johnston2016, Rossi2019, Braicovich2020, Li2020, Peng2020}.

Our results highlight the importance of developing a robust theoretical description of the complex \gls*{RIXS} process in order to analyse experimental data. The Green's function approach that we employ offers a practical way to incorporate the phonon and \gls*{e-p} interaction strength dispersions without the need for summation over all possible intermediate states. There are many avenues that remain to be explored, including the effects of itinerancy \cite{Bieniasz2020} and changes to the potential energy surface \cite{Geondzhian2018} in the intermediate state, and how the core hole affects the coupling probed by \gls*{RIXS} compared to other techniques \cite{Geondzhian2018}. The availability of theoretical methods to treat these effects will only become more pressing as the energy resolution of \gls*{RIXS} improves, and it is used to probe phonons in an ever increasing range of topical materials.

\begin{acknowledgments}
We thank Matteo Calandra for insightful discussions. C.D.D.~was supported by the Engineering and Physical Sciences Research Council (EPSRC) Centre for Doctoral Training in the Advanced Characterisation of Materials under Grant No.~EP/L015277/1. Work at Brookhaven National Laboratory was supported by the U.S.~Department of Energy, Office of Science, Office of Basic Energy Sciences, under Contract No.~DE-SC0012704. A.G.~was supported by a National Research Foundation of Korea grant funded by the Ministry of Science and ICT No.~2016K1A4A4A01922028, the U.S.~National Science Foundation under Grant No.~DMR-1904716, and acknowledges Research Computing at Arizona State University for providing high performance computing resources. K.G.~was supported by the U.S.~Department of Energy, Office of Science, Basic Energy Sciences as part of the Computational Materials Science Program. R.B.J.~acknowledges the support of the European Commission ``GREENDIAMOND'' project through the H2020 Large Project under Grant No.~SEP-210184415. Work at UCL was supported by the EPSRC under Grants No.~EP/N027671/1, EP/N034872/1, EP/H020055/1, and EP/N004159/1. We acknowledge SOLEIL for provision of synchrotron radiation facilities at the SEXTANTS beamline under proposals 20150221 and 20130732, and the Diamond Light Source for time on beamline I21 under proposal MM22695.
\end{acknowledgments}

\appendix

\section{\label{Exp_details}Experimental details}

Our carbon $K$-edge \gls*{RIXS} measurements were performed at beamline I21 of the Diamond Light Source. A natural graphite single crystal was mounted such that the scattering plane is $(h,0,l)$, and cleaved in vacuum. We chose an energy resolution of \SI{47}{\milli\electronvolt} (determined from scattering from amorphous carbon tape mounted next to the sample) to balance throughput and resolution. All \gls*{RIXS} spectra were taken with linear horizontal ($\pi$) incident x-ray polarisation, at a temperature of \SI{20}{\kelvin}, and with a counting time of \SI{20}{\minute}. X-ray absorption spectra in total electron yield were obtained by measuring the sample drain current. Data for energies around the $\pi^*$ resonance were taken with grazing-incident x-rays ($\theta =$ \ang{20}) while data around the $\sigma^*$ resonance were taken with normal-incident x-rays ($\theta =$ \ang{90}) in order to maximise the intensity of the phonon features. All data have been corrected for incident flux and self-absorption (see the Supplemental Material).

At these soft x-ray energies, the momentum transfer is confined to a small in-plane region of $< 0.1$ r.l.u.~around the zone center (see Fig.~\ref{RIXS_maps}c). As we used a fixed scattering angle $2\Theta =$ \ang{154}, the momentum transfer varies slightly with incident energy. Around the $\pi^*$ resonance the momentum transfer $\bm{q} = (0.0802 \pm 0.0004, 0, 0.164 \pm 0.001)$, while around the $\sigma^*$ resonance $\bm{q} = (0.0220 \pm 0.0002, 0, 0.300 \pm 0.002)$.

\section{\label{Theoretical_model}Theoretical model}

To model our rich experimental data we employed a phenomenological Green's-function formulation of the \gls*{RIXS} cross-section. In principle, final states involving both phonons and low-energy electronic excitations are allowed, the latter of which would result in a continuum of excitations. As such a continuum is not seen above the background in our data, however, we neglect these weak higher-order process and restrict the final states to those with only phonon excitations present.

We account for electron-electron interactions by constructing an effective quasi-particle exciton Green's function. Due to the different time scales, it is reasonable to separately consider electron-lattice interactions in the presence of an exciton which propagates through the material \cite{Antonius2017, Geondzhian2018}. Expanding the interacting exciton Green's function in a time series, we find two important types of diagrams: closed-loop diagrams where a phonon gets created and destroyed by the same exciton, and those where a phonon connects two excitons and gives rise to the final state phonon population. The former affect the intensities of the \gls*{RIXS} features non-linearly, and can be interpreted as contributions from intermediate state phonons. Evaluating these contributions has proven to be very computationally demanding in exact diagonalization approaches. Here, we account for them using the cumulant representation of the interacting Green's function, which reduces computational cost. The second type of diagram will be treated explicitly up to the number of phonon satellites visible in the experiment.

To simplify the notation, we omit electron-photon matrix elements which affect only the absolute intensity and not the characteristic shape of the phonon features, and drop the index of the excitonic state. The energy- and momentum-dependent cross-section then can be written as
\begin{equation}
    \sigma(\omega_i, \omega_f, \bm{q}) \approx - \sum_n |\Lambda^n(\omega_i, \bm{q})|^2 \Im D^n(\omega_i - \omega_f, \bm{q}) \label{cross-section}
\end{equation}
where $\Lambda^n(\omega, \bm{q})$ is the complex off-diagonal part of the exciton Green's function, representing the scattering of an exciton by $n$ phonons of total momentum $\bm{q} = \bm{k}_i - \bm{k}_f$, and the presence of phonons in the final state is reflected by the many-body Green's function $D^n(\omega, \bm{q})$. In the limit of infinite phonon life-time $\Im D^n(\omega, \bm{q}) = - \frac{1}{\pi} \delta(\omega - \sum_{i=1, n} \omega_i) \delta(\bm{q} - \sum_{i=1, n} \bm{q}_i)$. Here, we account for phonon-phonon interactions by attributing a small lifetime broadening to the phonon propagator.

Previous work has suggested an approximate form of $\Lambda$ in the limits of weak \cite{Devereaux2016} and intermediate \cite{Geondzhian2018} \gls*{e-p} interactions. For the purpose of this work, however, we use a closed-form solution for $\Lambda$ in the limit of local excitons coupled linearly to phonons. Although the cross-section in Eq.~(\ref{cross-section}) is given in the energy domain, from here on we use the time-dependent form. Focusing on the characteristic behavior of the phonon satellites with multiple momentum contributions, we have
\begin{equation}
    \Lambda^n(t, \bm{q}) = \Lambda^0(t) \prod_{\bm{q}_i...\bm{q}_n} \left[i G(\bm{q}_i) \int_0^t D^>(\tau, \bm{q}_i) d\tau \right] /\sqrt{n!}. \label{lambda}
\end{equation}
The diagonal part of the exciton Green's function is denoted $\Lambda^0$, and is the same dressed propagator that would appear in the absorption process ($I_\text{XAS} \sim \mathscr{F}\{\Lambda^0\}$). The bracketed part of Eq.~(\ref{lambda}) represents the phonon contributions to the scattering diagram, with the product over $n$ phonons conserving momentum, $\bm{q} = \bm{q}_1 + ... + \bm{q}_n$. $G(\bm{q})$ is the momentum-dependant electron-phonon interaction strength, and the greater phonon propagator in the zero-temperature limit is simply $D^>(t, \bm{q}) = -i \braket{b_{\bm{q}}(t) b_{\bm{q}}^\dagger(0)} = i \theta(t) e^{-i \omega(\bm{q}) t}$. Thus, the non-diagonal part of the exciton Green's function differs from the diagonal part by the interaction term with a certain number of final state phonons. Despite the simple form, in the limit of small excitons Eq.~(\ref{lambda}) accounts for all types of diagrams including those with vertex corrections.

The exciton propagator dressed by the exciton-phonon interaction can be expressed as $\Lambda^0(t) = L(t) e^{C(t)}$, where $L$ is the bare exciton Green's function. We accounted for the exciton lifetime by including a damping factor $e^{-\gamma t/2}$ in $L$. Note that this expression contains \textit{half} of the inverse lifetime, $\gamma / 2$. We used a value $\gamma / 2 =$ \SI{0.15}{\electronvolt}, slightly adjusted from the core-hole lifetime as determined by x-ray photoelectron spectroscopy measurements (\SI{0.1}{\electronvolt}) \cite{Sette1990} to account for the presence of the extra electron (see the Supplemental Material).

Finally, the cumulant function to second order in the exciton-phonon interaction can be written
\begin{equation}
    C(t) = [L(t)]^{-1}\int_0^t \int_0^t L(t-\tau_1) \Sigma_\text{FM}(\tau_1 - \tau_2) L(\tau_2) d\tau_1 d\tau_2 \label{cumulant}
\end{equation}
where the Fan-Migdal self-energy is $\Sigma_\text{FM} = -i \sum_{\bm{q}} G(\bm{q})^2 L(t) D(t, \bm{q})$. Substituting the exciton and phonon Green's functions, and taking the time integrals in Eq.~(\ref{cumulant}) analytically \cite{Langreth1970}, we get $C(t) = \sum_{\bm{q}} G(\bm{q})^2 / (\omega(\bm{q})^2 N) (e^{-i \omega(\bm{q}) t} + i \omega(\bm{q}) t - 1)$. It is important to notice that the summation runs over the whole Brillouin zone without restriction, since the total momentum transfer of the contributing diagrams is zero. From the form of Eqs.~(\ref{lambda}) and (\ref{cumulant}), it is clear that for any final state configuration there are contributions to the cross-section from phonons of all momenta. Therefore, even the intensity of the one-phonon peak depends on $G(\bm{q})$ throughout the Brillouin zone, not just at the experimental $\bm{q}$ point.

To evaluate Eq.~(\ref{cumulant}), we used the TO phonon dispersion calculated with density functional perturbation theory. Brillouin zone integrations were performed using a reduced set of weighted momentum points which reflect the symmetries of the graphite structure \cite{Monkhorst1976}. Having calculated all the contributions to Eq.~(\ref{lambda}) in the time domain, we can then perform a Fourier transform at the given incident photon energy and plug it into Eq.~(\ref{cross-section}) to obtain the \gls*{RIXS} cross-section.

\bibliography{ref}

\end{document}

\title{Supplemental Material for ``Probing electron-phonon interactions away from the Fermi level with resonant inelastic x-ray scattering''}

\author{C.~D.~Dashwood}\email{cameron.dashwood.17@ucl.ac.uk}
\thanks{These authors contributed equally}
\affiliation{London Centre for Nanotechnology and Department of Physics and Astronomy, University College London, London, WC1E 6BT, UK}

\author{A.~Geondzhian}
\thanks{These authors contributed equally}
\affiliation{Max Planck POSTECH/KOREA Research Initiative, 37673 Pohang, South Korea}
\affiliation{Max Planck Institute for the Structure and Dynamics of Matter, Luruper Chaussee 149, 22761 Hamburg, Germany}

\author{J.~G.~Vale}
\affiliation{London Centre for Nanotechnology and Department of Physics and Astronomy, University College London, London, WC1E 6BT, UK}

\author{A.~C.~Pakpour-Tabrizi}
\affiliation{London Centre for Nanotechnology and Department of Electronic and Electrical Engineering, University College London, London, WC1E 6BT, UK}

\author{C.~A.~Howard}
\author{Q.~Faure}
\author{L.~S.~I.~Veiga}
\affiliation{London Centre for Nanotechnology and Department of Physics and Astronomy, University College London, London, WC1E 6BT, UK}

\author{D.~Meyers}
\affiliation{Condensed Matter Physics and Materials Science Department, Brookhaven National Laboratory, Upton, New York 11973, USA}
\affiliation{Department of Physics, Oklahoma State University, Stillwater, Oklahoma 74078, USA}

\author{S.~G.~Chiuzb\u{a}ian}
\affiliation{Synchrotron SOLEIL, L'Orme des Merisiers, Saint-Aubin, BP 48, 91192 Gif-sur-Yvette, France}
\affiliation{Sorbonne Universit\'{e}, CNRS, Laboratoire de Chimie Physique-Mati\'{e}re et Rayonnement, UMR 7614, 4 place Jussieu, 75252 Paris Cedex 05, France}

\author{A.~Nicolaou}
\author{N.~Jaouen}
\affiliation{Synchrotron SOLEIL, L'Orme des Merisiers, Saint-Aubin, BP 48, 91192 Gif-sur-Yvette, France}

\author{R.~B.~Jackman}
\affiliation{London Centre for Nanotechnology and Department of Electronic and Electrical Engineering, University College London, London, WC1E 6BT, UK}

\author{A.~Nag}
\author{M.~Garc\'{i}a-Fern\'{a}ndez}
\author{Ke-Jin Zhou}
\author{A.~C.~Walters}
\affiliation{Diamond Light Source, Didcot, Oxfordshire, OX11 0DE, UK}

\author{K.~Gilmore}
\affiliation{Condensed Matter Physics and Materials Science Department, Brookhaven National Laboratory, Upton, New York 11973, USA}
\affiliation{Physics Department and IRIS Adlershof, Humboldt-Universit\"{a}t zu Berlin, Zum Gro{\ss}en Windkanal 2, 12489 Berlin, Germany}
\affiliation{European Theoretical Spectroscopy Facility (ETSF)}

\author{D.~F.~McMorrow}
\affiliation{London Centre for Nanotechnology and Department of Physics and Astronomy, University College London, London, WC1E 6BT, UK}

\author{M.~P.~M.~Dean}\email{mdean@bnl.gov}
\affiliation{Condensed Matter Physics and Materials Science Department, Brookhaven National Laboratory, Upton, New York 11973, USA}

\maketitle

\section{Self-absorption correction}

Self-absorption -- the re-absorption by the sample of previously scattered photons -- depends on the energies and angles of both the incident and outgoing x-rays, and must therefore be corrected to reliably compare spectra. We follow previous \gls*{RIXS} studies \cite{Rossi2019, Vale2019, Feng2020} and define a self-absorption correction factor
\begin{equation}
    \label{corr_factor}
    C_{\bm{\epsilon}, \bm{\epsilon}'}(2\Theta, \theta, \omega_i, \omega_o) = \frac{1}{1 + u(2\Theta, \theta) t_{\bm{\epsilon}, \bm{\epsilon}'}(\omega_i, \omega_o)}.
\end{equation}
In this expression, $u = \sin{\theta} / \sin{(2\Theta - \theta)}$ is a geometrical term that depends on the scattering angle $2\Theta =$ \ang{154}, and sample angle $\theta =$ \ang{20} (\ang{90}) for the $\pi^*$ ($\sigma^*$) spectra. The other term
\begin{equation}
    t_{\bm{\epsilon}, \bm{\epsilon}'}(\omega_i, \omega_o) = \frac{\alpha_0 + \alpha_{\bm{\epsilon^\prime}}(\omega_o)}{\alpha_0 + \alpha_{\bm{\epsilon}}(\omega_i)}
\end{equation}
depends on $\alpha_0$, the absorption coefficient before the resonance (taken at \SI{284}{\electronvolt} for the $\pi^*$ resonance and \SI{289}{\electronvolt} for the $\sigma^*$ resonance), and $\alpha_{\bm{\epsilon}}(\omega)$, the resonant absorption coefficient for x-rays with energy $\omega$ and polarization $\bm{\epsilon}$.

\begin{figure*}
	\includegraphics[width=\linewidth]{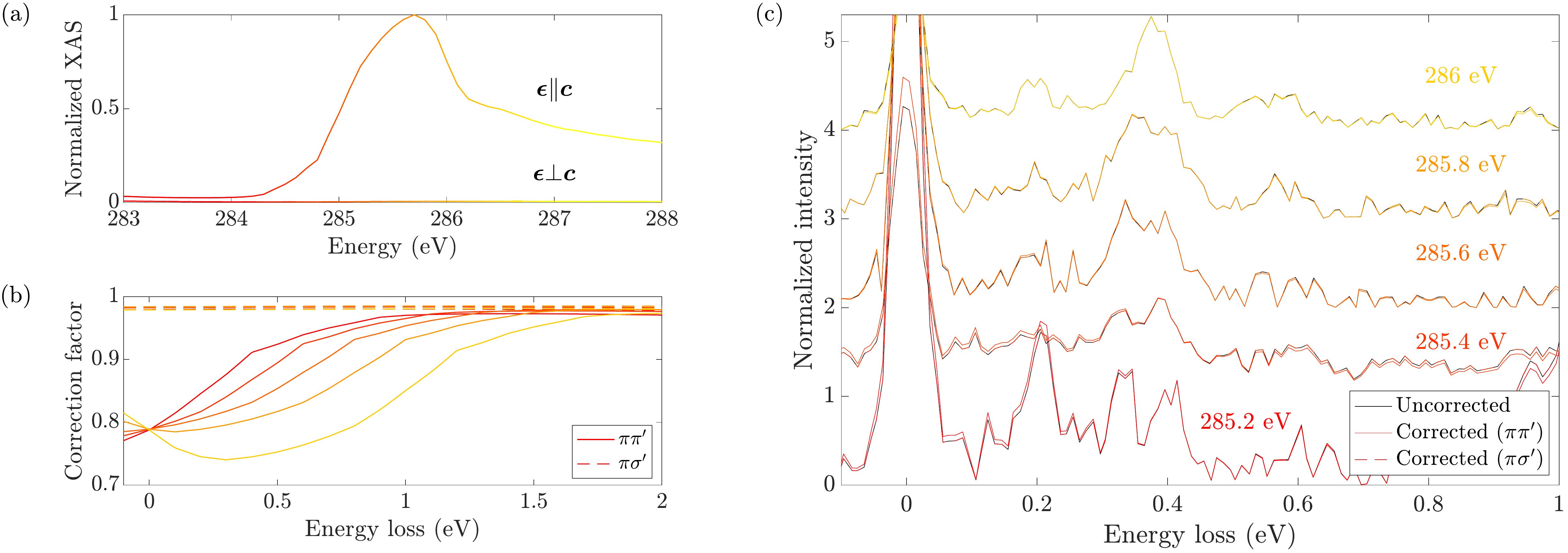}
	\caption{\label{pi_abs_corr}(a) \gls*{XAS} around the $\pi^*$ resonance with incident x-rays polarized parallel ($\bm{\epsilon}\|\bm{c}$) and perpendicular ($\bm{\epsilon}\perp\bm{c}$) to the $\bm{c}$ axis. (b) Self-absorption correction factors in the $\pi\pi'$ (solid) and $\pi\sigma'$ (dashed) polarization channels, with colours corresponding to the incident energies labelled in (c). (c) Uncorrected (black) and corrected (red to yellow) \gls*{RIXS} spectra in the two polarization channels, normalized to the intensity of the two-phonon peak. The corrections introduce changes below the noise level to the normalized spectra.}
\end{figure*}

\begin{figure*}
	\includegraphics[width=\linewidth]{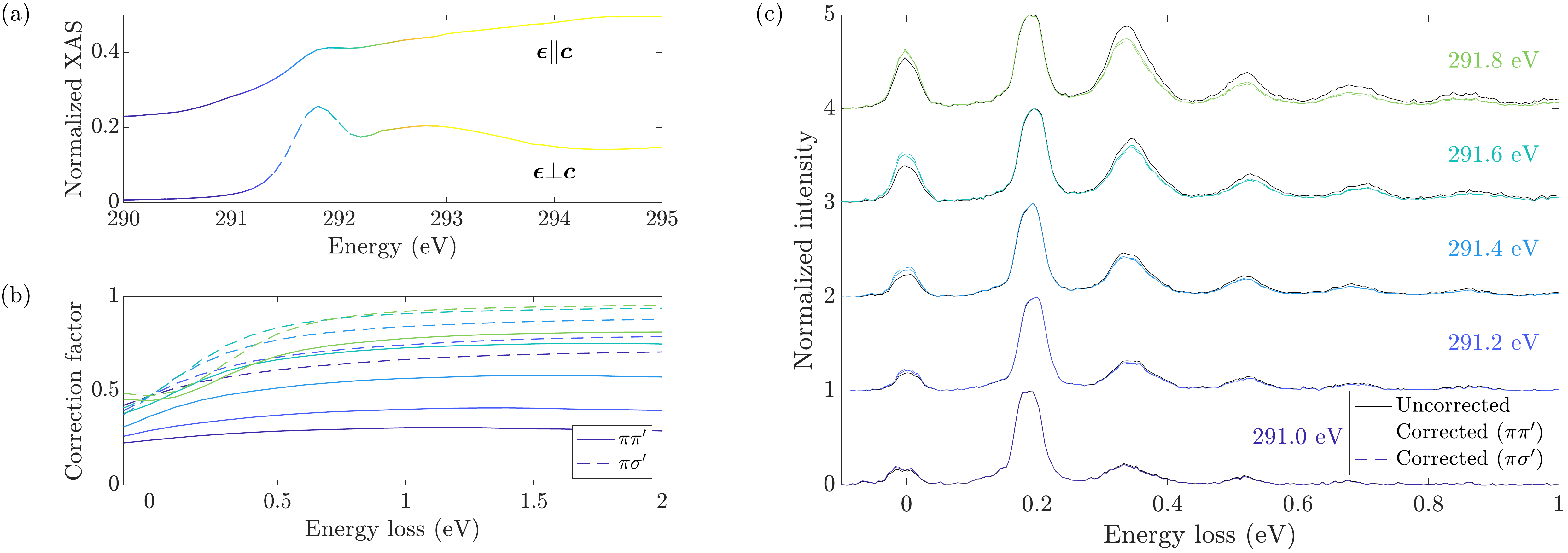}
	\caption{\label{sig_abs_corr}(a) \gls*{XAS} around the $\sigma^*$ resonance with incident x-rays polarized parallel ($\bm{\epsilon}\|\bm{c}$) and perpendicular ($\bm{\epsilon}\perp\bm{c}$) to the $\bm{c}$ axis. In this work we focus on the $\sigma^{*1}$ resonance at \SI{291.8}{\electronvolt} and ignore the broad $\sigma^{*2}$ resonance above which arises from zero-point vibrations. (b) Self-absorption correction factors in the $\pi\pi'$ (solid) and $\pi\sigma'$ (dashed) polarization channels, with colours corresponding to the incident energies labelled in (c). (c) Uncorrected (black) and corrected (blue to green) \gls*{RIXS} spectra in the two polarization channels, normalized to the intensity of the one-phonon peak.}
\end{figure*}

The absorption coefficients can be obtained from \gls*{XAS} measured in total electron yield, shown in Fig.~\ref{pi_abs_corr}(a) and \ref{sig_abs_corr}(a). With our incident x-rays polarized in the $(h, 0, l)$ scattering plane ($\pi$ polarization), we measure the absorption for $\bm{\epsilon}\|\bm{c}$ ($\bm{\epsilon}\perp\bm{c}$) using x-rays incident normal (grazing) to the sample surface. From these, we calculate the absorption coefficients as
\begin{align}
    \alpha_\pi(\omega_i) &= \text{XAS}(\omega_i, \bm{\epsilon}\perp\bm{c}) \sin^2{\theta} + \text{XAS}(\omega_i, \bm{\epsilon}\|\bm{c}) \cos^2{\theta}, \nonumber \\
    \alpha_\pi(\omega_o) &= \text{XAS}(\omega_o, \bm{\epsilon}\perp\bm{c}) \sin^2{(2\Theta - \theta)} + \text{XAS}(\omega_o, \bm{\epsilon}\|\bm{c}) \cos^2{(2\Theta - \theta)}, \nonumber \\
    \alpha_\sigma(\omega_o) &= \text{XAS}(\omega_o, \bm{\epsilon} \perp \bm{c}).
\end{align}
The resulting correction factors are shown in Fig.~\ref{pi_abs_corr}(b) and \ref{sig_abs_corr}(b). The \gls*{RIXS} spectra are then corrected by dividing through by $C_{\bm{\epsilon}, \bm{\epsilon}'}(2\Theta, \theta, \omega_i, \omega_o)$.

The anisotropic absorption of graphite results in a significant polarization dependence of the correction factors, especially around the $\pi^*$ resonance. The polarization of the outgoing photons will depend on the symmetry of the phonon modes being excited, and was not determined in our experiment. As our model only predicts the relative intensities of the phonon peaks, however, we can normalize the spectra to the intensity of the strongest multi-phonon peak. As shown in Fig.~\ref{pi_abs_corr}(c) and \ref{sig_abs_corr}(c), after normalisation the difference between the corrected spectra is minimal over the energy range of interest, and does not significantly affect the fits to our model. We therefore correct all spectra assuming that the outgoing photons are $\pi$ polarised.

\section{Assignment of phonon modes}

\begin{figure}
	\includegraphics[width=\linewidth]{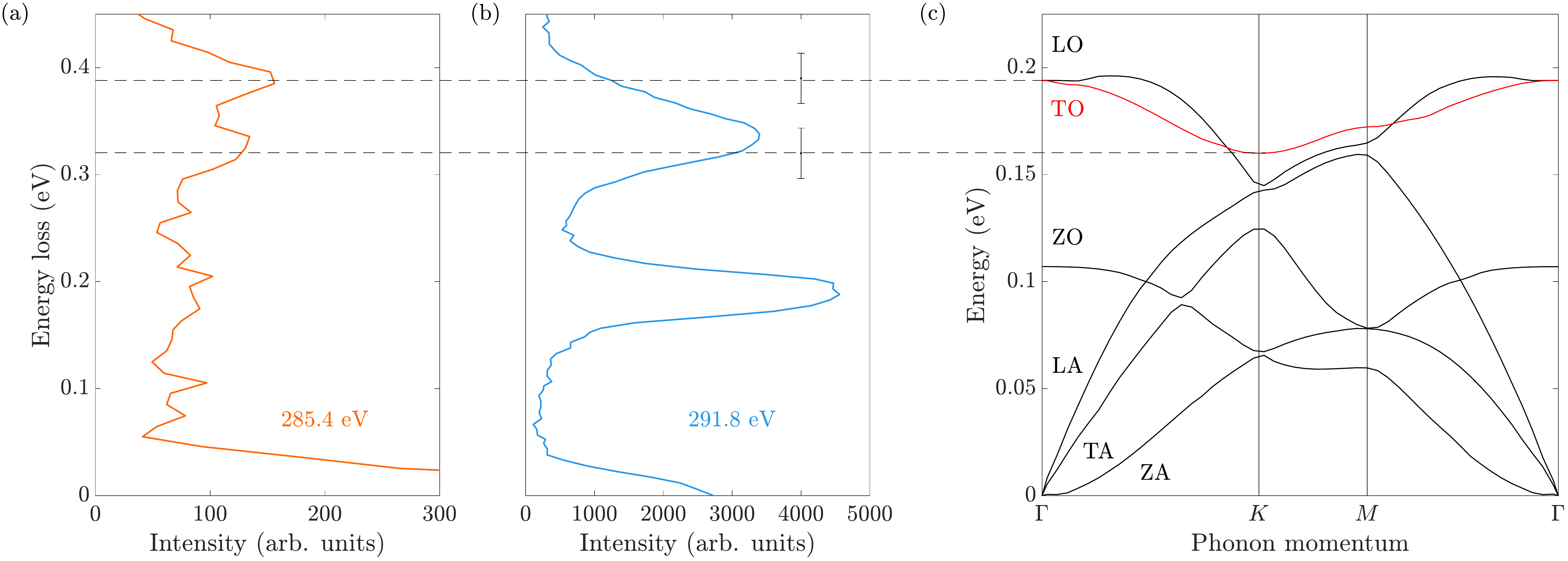}
	\caption{\label{Mode_assignment}Experimental \gls*{RIXS} spectra at (a) \SI{285.4}{\electronvolt} and (b) \SI{291.8}{\electronvolt}, alongside (c) the phonon dispersion of graphite from Ref.~\cite{Mohr2007} scaled to half the energy range of (a) and (b). The horizontal dashed lines mark (twice) the energies of the TO mode at $\Gamma$ and $K$, and the vertical error bars show the instrumental energy resolution.}
\end{figure}

In the main text, we assign the contributions to the \gls*{RIXS} spectra based on the known phonon dispersion of graphite \cite{Mohr2007}. Here we demonstrate this process for the two-phonon features. Fig.~\ref{Mode_assignment} shows spectra near the $\pi^*$ and $\sigma^*$ resonances plotted alongside the phonon dispersion, scaled such that the phonon energies line up with twice the energy loss of the spectra. From this, it is clear that the two peaks in the split feature of the $\pi^*$ spectrum occur at twice the energies of transverse optical (TO) phonons at $\Gamma$ and $K$, while the broad asymmetric feature of the $\sigma^*$ spectrum spans twice the bandwidth of the TO mode. Taking into account the \SI{47}{\milli\electronvolt} experimental energy resolution, it can be seen that contributions from the acoustic (ZA, TA, and LA) and out-of-plane optical (ZO) modes are negligible, and the longitudinal optical (LO) mode could only have significant contributions when it is nearly degenerate with the TO mode.

\section{Intermediate-state lifetime}

\begin{figure}
	\includegraphics[width=\linewidth]{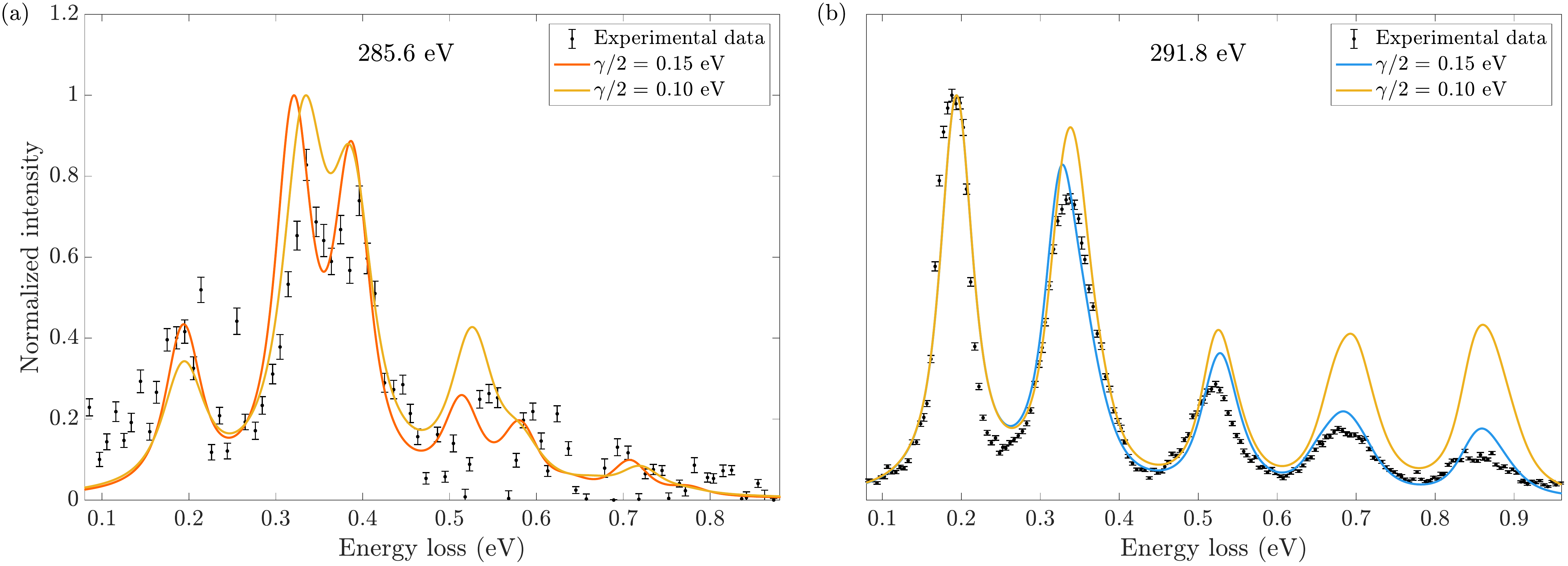}
	\caption{\label{Lifetime_comp}Comparison of best-fit \gls*{RIXS} spectra at (a) the $\pi^*$ and (b) the $\sigma^*$ resonance with half inverse intermediate-state lifetimes of $\gamma / 2 =$ \SI{0.15}{\electronvolt} (orange/blue) and \SI{0.10}{\electronvolt} (yellow).}
\end{figure}

To model our \gls*{RIXS} spectra we adjusted the intermediate-state lifetime from the value $\gamma / 2 =$ \SI{0.1}{\electronvolt} (half-width at half-maximum of the Lorentzian component of the spectral peak) measured by x-ray photoemission spectroscopy \cite{Sette1990} due to the presence of the extra excited electron in \gls*{RIXS}. Figure \ref{Lifetime_comp} compares best-fit \gls*{RIXS} spectra at the $\pi^*$ and $\sigma^*$ resonances calculated using the bare core-hole lifetime \SI{0.1}{\electronvolt}, to the adjusted value \SI{0.15}{\electronvolt}. We can see that the adjusted lifetime gives better agreement with both experimental spectra, especially for the high-order peaks at the $\sigma^*$ resonance. This adjusted lifetime was used for all calculations shown in the main manuscript, successfully reproducing the energy dependence of the experimental spectra below both resonances.

\section{Independent fitting of the energy-dependent spectra}

\begin{figure}
	\includegraphics[width=\linewidth]{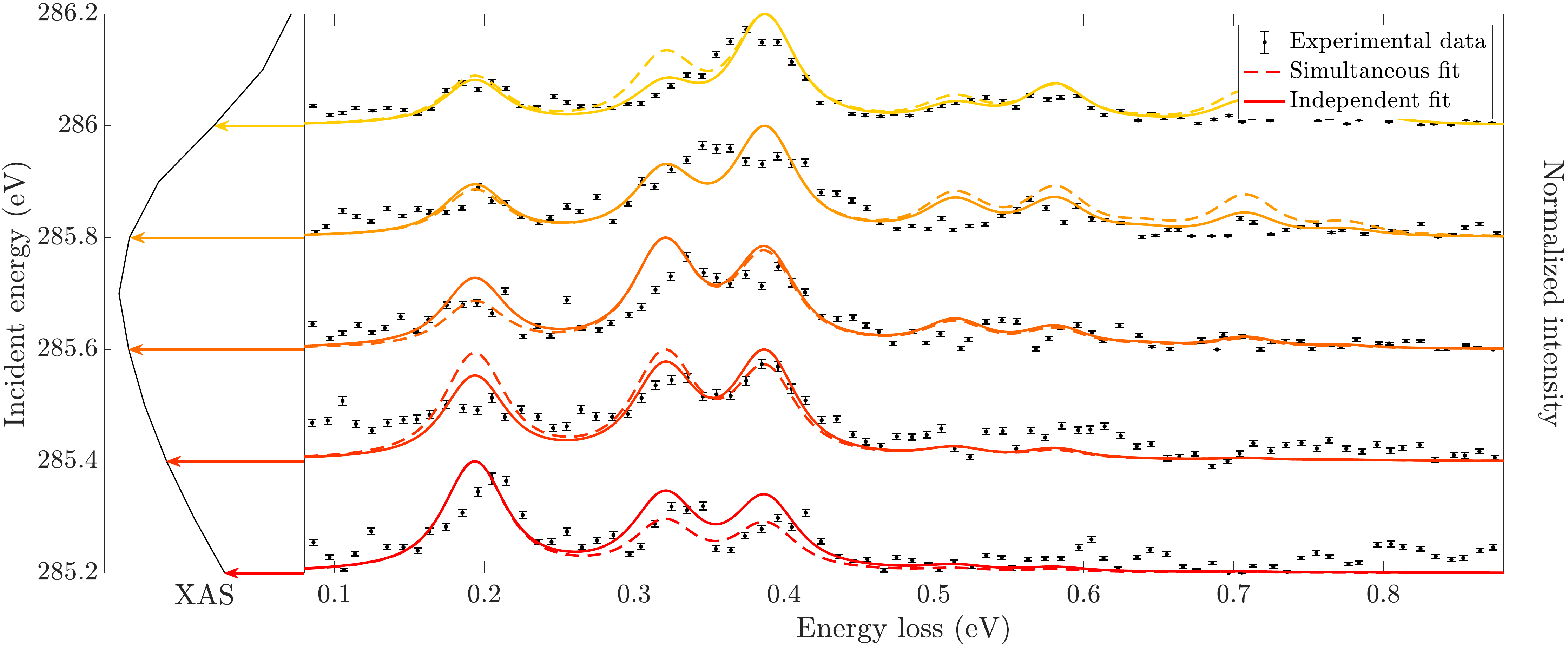}
	\caption{\label{Independent_fit}Comparison of independent (solid colored lines) and simultaneous (dashed coloured lines) fits to the experimental \gls*{RIXS} spectra (black points with error bars) around the $\pi^*$ resonance, plotted alongside the \gls*{XAS} with arrows indicating the energies.}
\end{figure}

The calculated spectra shown in Fig.~3(c) of the main manuscript are the result of a simultaneous fit to the experimental spectra around the $\pi^*$ resonance, such that the only parameter of the model that is varied between the spectra is the incident energy.

We have also fitted the spectra at each energy independently to see how this affects the extracted \gls*{e-p} interaction strengths. Figure \ref{Independent_fit} shows a comparison of the simultaneous and independent fits, with the resulting values of $G(\bm{q})$ listed in Table \ref{Interaction_strengths}. Both sets of calculated spectra have an acceptable level of agreement with the experimental data given the noise level, but the independent fits do offer better agreement especially for the spectrum at \SI{286}{\electronvolt}. As we note in the main text, we expect composition of the intermediate electronic state and therefore the momentum dependence of $G$ to vary in the high-energy tail of the \gls*{XAS} peak, providing a natural explanation for this. While Table \ref{Interaction_strengths} shows some variation in the extracted interaction strengths, the general trend of $G(\bm{q})$ through the Brillouin zone -- with $G(K) > G(\Gamma) >> G(K - \bm{q}_\text{exp}) \geq G(\bm{q}_\text{exp})$ -- is valid irrespective of the fitting procedure.

\begin{table}
    \caption{\label{Interaction_strengths}$G(\bm{q})$ extracted from simultaneous and independent fits to the \gls*{RIXS} spectra around the $\pi^*$ resonance, at the momenta $\Gamma$, $\bm{q}_\text{exp} \approx (0.08, 0.00)$, $K - \bm{q}_\text{exp}$, and $K$.}
    \begin{ruledtabular}
        \begin{tabular}{c|cccc}
            $G$ (\SI{}{\electronvolt}) & $\bm{q} = \Gamma$ & $\bm{q} = \bm{q}_\text{exp}$ & $\bm{q} = K - \bm{q}_\text{exp}$ & $\bm{q} = K$ \\
            \hline
            Simultaneous fit & 0.23 & 0.05 & 0.06 & 0.32 \\
            $\hbar \omega_i =$ \SI{285.2}{\electronvolt} & 0.24 & 0.05 & 0.13 & 0.30 \\
            $\hbar \omega_i =$ \SI{285.4}{\electronvolt} & 0.24 & 0.05 & 0.07 & 0.30 \\
            $\hbar \omega_i =$ \SI{285.6}{\electronvolt} & 0.23 & 0.07 & 0.07 & 0.32 \\
            $\hbar \omega_i =$ \SI{285.8}{\electronvolt} & 0.20 & 0.05 & 0.05 & 0.28 \\
            $\hbar \omega_i =$ \SI{286.0}{\electronvolt} & 0.24 & 0.05 & 0.12 & 0.27 \\
        \end{tabular}
    \end{ruledtabular}
\end{table}

\section{Comparison with previous RIXS measurements on graphite}

Feng \textit{et al.} recently presented lower-resolution \gls*{RIXS} measurements around the $\sigma^*$ resonance of graphite \cite{Feng2020}, interpreting their data with reference to Raman spectroscopy results. Their analysis implicitly assumed, however, a coupling of phonons to electrons near the Fermi level, leading them to constrain $G(\bm{q})$ to be non-zero only at the $\Gamma$ and $K$ points. While this is reasonably accurate for the $\pi^*$ resonance, we show in the main text that \gls*{RIXS} at the $\sigma^*$ resonance probes the interaction of phonons with elections \textit{away} from the Fermi level, whose strength is significant throughout the Brillouin zone. As well as overlooking the possibilities afforded by tuning the incident photon energy, this serves as a warning that careful consideration of the states involved in the \gls*{RIXS} process is essential for correct interpretation of experimental spectra, especially if the energy resolution is lacking.

\begin{figure}
	\includegraphics[width=\linewidth]{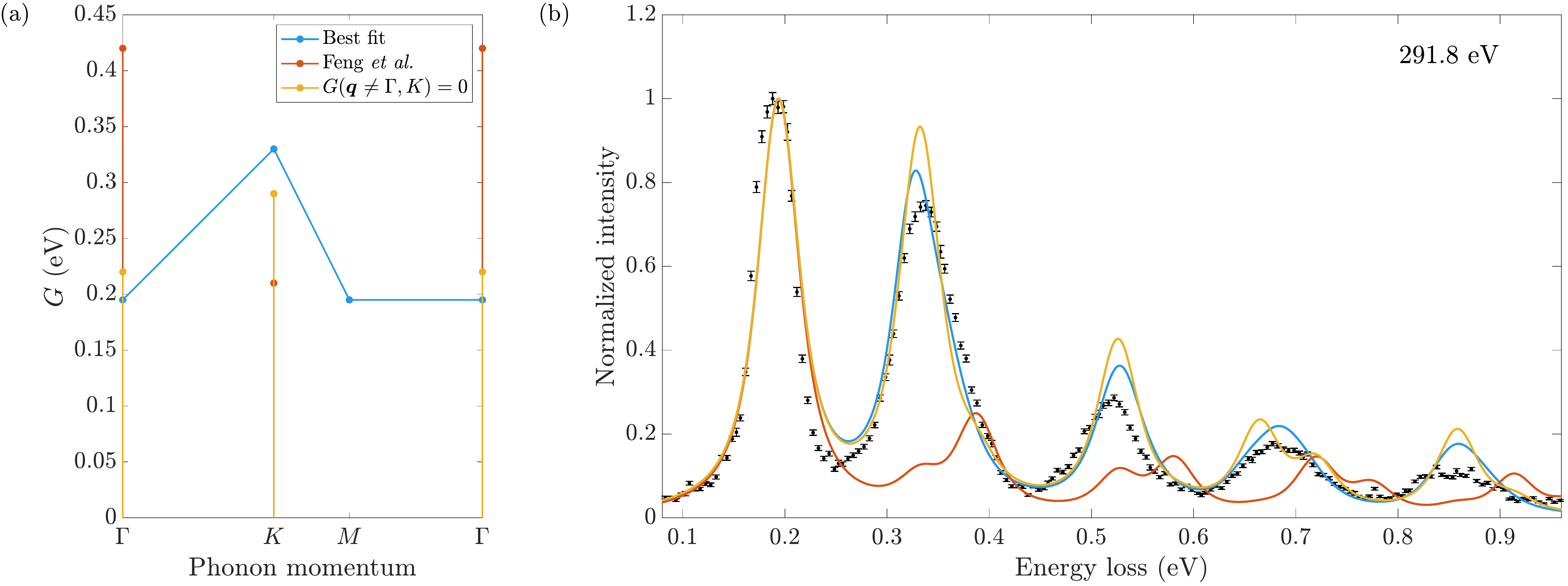}
	\caption{\label{LBNL_comp}(a) Our best-fit $G(\bm{q})$ for $\sigma^*$ electrons (blue), compared to that obtained by Feng \textit{et al.} \cite{Feng2020} (orange) and our best-fit with $G$ constrained to be finite only at $\bm{q} = \Gamma$ and $K$ (yellow). (b) Comparison of calculated \gls*{RIXS} spectra at the $\sigma^*$ resonance, with colours corresponding to those in (a) and the experimental spectrum shown as black points with error bars. All spectra are calculated with an intermediate-state lifetime $\gamma / 2 =$ \SI{0.15}{\electronvolt}.}
\end{figure}

We compare our best-fit $G(\bm{q})$ at the $\sigma^*$ resonance (blue), to that obtained by Feng \textit{et al.} (orange) in Fig.~\ref{LBNL_comp}(a). From the resulting normalized spectra in Fig.~\ref{LBNL_comp}(b), we see that the couplings from Feng \textit{et al.} give a poor fit to all of the multi-phonon peaks.

We can also investigate the effect of phonon mode mixing in the intermediate \gls*{RIXS} state through a direct comparison to the results of Feng \textit{et al.}, who neglect it in their modeling. Even if the final \gls*{RIXS} state consists of only a single phonon, the intermediate state can contain many modes which non-linearly affect the cross-section \cite{Geondzhian2020a}. Let us briefly set aside what we know about the momentum dependence of $G$ away from the Fermi level, and fit our $\sigma^*$ spectrum with contributions from only $\Gamma$ and $K$ (yellow in Fig.~\ref{LBNL_comp}). While the calculated spectrum does not agree with the experimental data as well as our best fit, especially in the high-order peaks which become split, we see that the values of $M_\Gamma$ and $M_K$ approach those of our linearly-varying model. Our model yields a ratio $G(\Gamma) / G(K) = 0.79$, compared to $G(\Gamma) / G(K) = 2.1$ obtained by Feng \textit{et al.} (note that $G(\Gamma) / G(K)$ is insensitive to the different intermediate-state lifetimes used, as in the Ament model the relative peak intensities depend on the ratio of $G$ to lifetime). We therefore see that, even for the relatively flat optical modes studied here, accounting for intermediate-state mixing strongly affects the momentum-dependence of $G$ extracted from the data.

\section{Momentum dependence of the RIXS spectra}

\begin{figure}
	\includegraphics[width=0.8\linewidth]{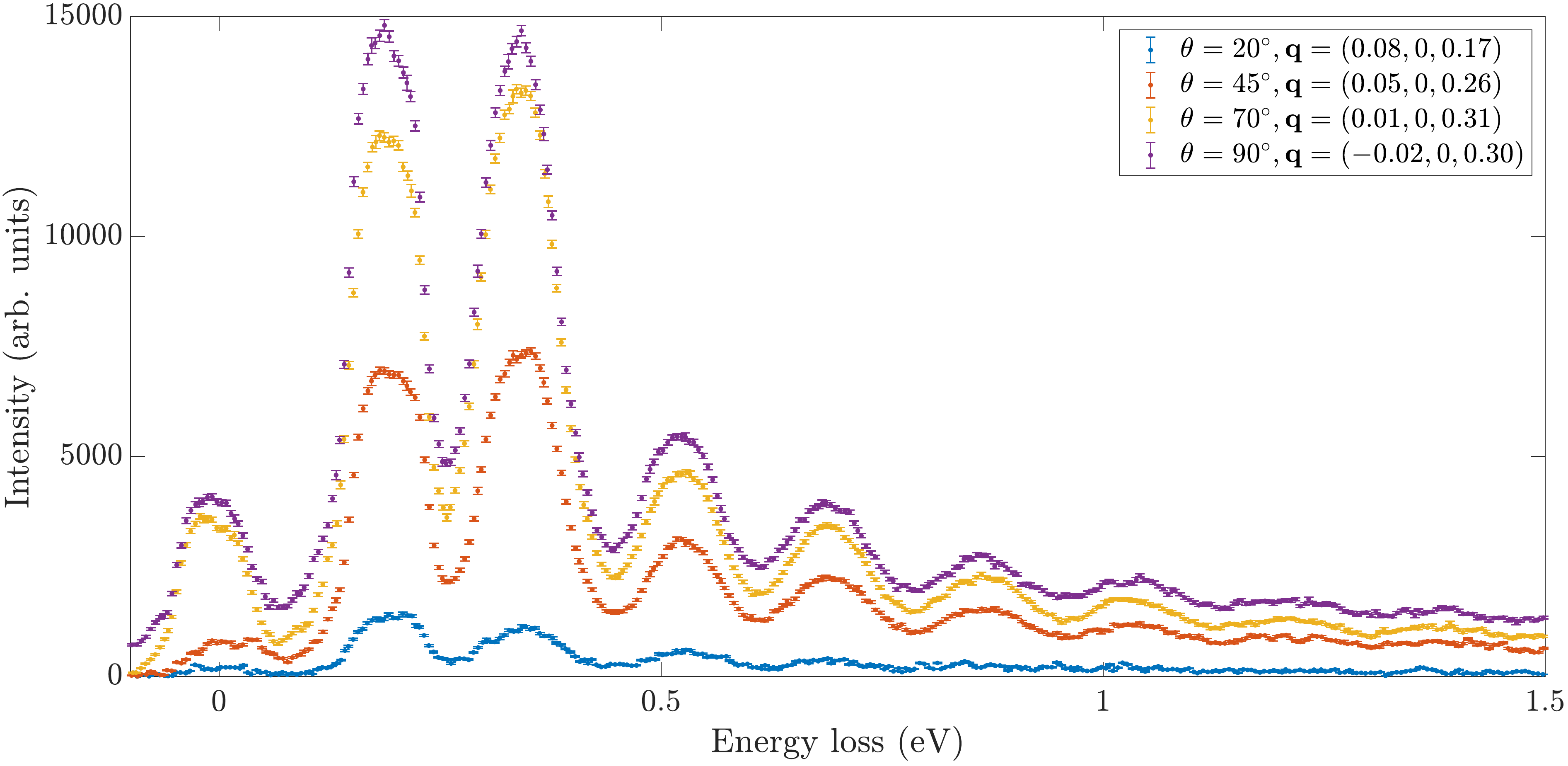}
	\caption{\label{Momentum_dep}Momentum dependence along $(h, 0, l)$ of the experimental \gls*{RIXS} spectra at the $\sigma^*$ resonance.}
\end{figure}

The maximum momentum transfer is limited by the low energy of the carbon $K$ edge. Additionally, the \gls*{RIXS} cross-section falls off rapidly for sample angles away from $\theta =$ \ang{0} (grazing incidence) at the $\pi^*$ resonance, and away from $\theta =$ \ang{90} (normal incidence) at the $\sigma^*$ resonance. In the main manuscript, we show how these restrictions can be circumvented by analysing the contributions to the multi-phonon peaks, which can come from phonons throughout the Brillouin zone.

Figure \ref{Momentum_dep} shows the momentum-dependence of the \gls*{RIXS} spectra at the $\sigma^*$ resonance along the $(h, 0, l)$ direction. The in-plane momentum was varied by changing the sample angle $\theta$ while keeping the spectrometer angle $2\Theta (\neq 2 \times \theta)$ fixed, so the total momentum transfer $q \propto \sqrt{4h^2 / (3a^2) + l^2 / c^2}$ is constant. While the overall intensity falls off rapidly with decreasing $\theta$, it can be seen that the relative intensities of the phonon peaks change only minimally, supporting our conclusion that $G(\bm{q})$ varies gradually away from the $\Gamma$ point.

The low intensity of the phonon features at the $\pi^*$ resonance, along with the strong fall-off in intensity with increasing $\theta$, precluded measurement of a momentum-dependence with reasonable statistics.

\bibliography{ref}